\documentstyle[preprint,aps,prl]{revtex}
\tightenlines

\begin{document}
\draft
\pagestyle{headings}
\preprint{yan.tex}
\title{Instability and Periodic Deformation in Bilayer Membranes 
Induced by Freezing}

\author{Yan Jie$^{1}$, Zhou    Haijun$^{1}$,  Ou-Yang      Zhong-can$^{1,2}$}
\address{
$^{1}$Institute of Theoretical Physics, Chinese Academy of Sciences,\\ 
P.O. Box 2735, Beijing 100080, China\\
$^{2}$Center for Advanced Study, Tsinghua University, Beijing 100084, China}
\date{\today}

\maketitle
\begin{abstract}
The instability and periodic deformation of bilayer membranes during freezing
processes 
 are studied as a function of
 the  difference of  the shape energy 
between the  high and the low temperature membrane states. 
It is shown that there exists a threshold stability condition, bellow which a planar configuration will be deformed. Among the deformed shapes,
the periodic curved square textures  are shown being one kind of the
solutions of the associated shape equation. In consistency with recent experimental observations, the optimal ratio
of period and amplitude for such a texture  is found to be approximately
equal to $\sqrt{2}\pi$. 
\end{abstract}
\pacs{PACS numbers:87. 22. Bt,   82. 70. -y, 82. 65. -i}

\narrowtext
Lipid molecules, such as phospholipids, assemble into bilayer 
membranes (BMs) in aqueous environment. On the elasticity and statistics of closed
 BMs (vesicles), many investigations have been 
taken on[1-4].
 At high enough concentration of lipids the BMs  will convert from vesicles into
 extended configurations[5]. For extended membranes, experiment
demonstrated the existence of one-dimensional periodic cylindrical surfaces[6].
It is found theoretically that such kind of shapes 
can be well explained by the Helfrich curvature elasticity theory for BMs[7].
Recent experiments employing freeze fracture electron
 microscopy method
revealed the existence of stable periodic curved square textures 
(PCSTs), which resemble egg
cartons and look equal from both sides of the membrane[8-10]. 
 Later on, Kl{\"{o}}sgen  and Helfrich
found a grainy texture in vesicular bilayers of egg yolk phosphatidylcholine 
with cryo-transmission electron microscopy[11].  In fact, periodic
deformations of cell membrane are abundant in nature[12], and
  a common aspect of these observations[8-12]
 is that all these curved textures are created through 
rapid cooling processes. 

On the theoretical side, Meyer conjected   
the observed periodic shape to be 
infinite periodic minimal surface (IPMS)[13].  As shown in
Ref. [14], the formation of IPMS requires the regular arrangement
of proteins or other globular macromolecules, however, this condition
is not necessary  for the formation of the PCSTs reported in
Refs. [8] and [9]. 
Recently, an approach for the problem is proposed by 
Goetz and Helfrich by suggesting a new curvature elasticity model in which bending energy terms higher than quadratic order in the principal 
curvatures are included and used to Monte Carlo simulation(MCS) for yielding PCSTs[15]. 
 They found that if the MCS is carried out initially from a plane, then the plane
 remains unchanged apart from its fluctuations, which indicates the existence of
  an energy barrier
 between a planar and any periodicly curved shapes[15].   As an analytic check for
 the problem, our first task in this Letter is to show  that if taking account only
  curvature elastic energy (no matter whether 
higher order energy terms be included or not), a
plane will always be a stable configuration and an energy barrier must be overcome if some possible stable curved
pattern is to be realized. Therefore, the question on the
 origin of the formation of PCSTs is
still open. As the second step, the main propose in this Letter, the PCST formation is considered
as the result of the quench of the temperature decreasing.  
We analytically derived the  equilibrium-shape equation for the BMs
by taking account of competition among
the curvature elasticity, the   volume Gibbs free energy difference
 between the high  and the low temperature membrane states, 
 and the surface tension energy difference. 
The sum of these three energies can be understood as the shape 
formation energy (see bellow for details).
The  planar configuration is a 
trivial solution of the shape equation, and for small violation from a plane,
 a PCST described by triangular functions is also another (approximate) solution 
if the differences of Gibbs free energy density and tension density 
satisfy some threshold condition. 
We show that below such a threshold condition
 the shape formation energy of a planar BM in the quench process
could become negative. In other words, the planar BM
becomes unstable in the freezing process, as a result,  
curved BMs will be formed spontaneously to keep the equilibrium condition 
($i.e.$ the zero shape formation energy).
Taking into consideration of the equilibrium condition in the 
quench-like cooling processes, the above argument provides an insight 
for the mechanism of the PCST  deformation for the BMs.
The optimal ratio of the periods and amplitudes of the PCSTs 
 formed in 
such processes is shown  to be approximately equal to $\sqrt{2}\pi $. 
This result 
is  well consistent with recent 
observations of the PCST formation in BMs by several groups[8-10]. 

First, we check whether the instability of a planar configuration is caused by
higher order energy terms or not. 
 We follow the work of Goetz and Helfrich[15] and express the bending
energy as
\begin{equation}
F_{B}=\int [ {\frac{1}{2}} \kappa H^{2}+{\overline{\kappa}}K+
\kappa_{2} K^{2}+\kappa_{4} K^{4} ] dA,
\end{equation}
where $H=(c_{1}+c_{2})/2$ and  $K=c_{1} c_{2}$  are, respectively,  the
mean curvature and the Gaussian curvature of the membrane surface,
$ \kappa$ is  the bending rigidity and ${\overline{\kappa}}$ is the 
modulus of Gaussian curvature, $dA$ is the surface area element, $\kappa_{2}$
is supposed to be negative[15]. Considering periodic deformations of the
flat membrane, we omit the ${\overline{\kappa}}$ term because of the
Gauss-Bonnet theorem. The general form of a surface in the Cartesian
coordinate system is described as
\begin{equation}
{\bf Y}(x, y)=( x, y, z( x, y ) ).
\end{equation}
The corresponding equilibrium-shape equation for the surface described by
Eq. (2)  is obtained by requiring the first order variation of its elastic
energy (1) with small deformations, $\delta^{(1)} F_{B}$, equal to zero.
This condition is satisfied if[16, 17]
\begin{equation}
\kappa H^{3}-\kappa H K + 2 \kappa_{2} H K^{2}+6 \kappa_{4} H K^{4}+
{\frac{1}{2}}\kappa {\nabla}^2 H +2\kappa_{2} \overline{\nabla}^{2}K +4 \kappa_{4}
\overline{\nabla}^{2} K^{3} =0,
\end{equation}
where $\nabla^{2}=(1/\sqrt{g})\partial_{i}(\sqrt{g} g^{ij}\partial_{j})$
 is the Laplace-Beltrami operator,  and 
$\overline{\nabla}^{2}$ is a new operator defined as $(1/\sqrt{g})\partial_{i}(
\sqrt{g} K L^{ij}\partial_{j})$, here ${\partial}_1={\partial}_x$, 
${\partial}_2={\partial}_y$, $g=det(g_{ij})$, $(L^{ij})=(L_{ij})^{-1}$, and 
$g_{ij}$, $L_{ij}$ are associated with the first and second fundamental
 forms of the surface, respectively
[16, 17].
Eq. (3) is a highly nonlinear differential equation and  generally
it is very difficult to find exact solutions. The egg carton surfaces
obtained by Goetz and Helfrich[15] may be considered as a numerical periodic solution
 of Eq. (3). Although the planar configuration, such as ${\bf Y}= ( x, y, 0 )$ with
  its $H = K =0$, 
 is a exact solution of it, 
the second variation of $F_{B}$ for  the plane has been calculated as
\begin{equation}
\delta^{(2)} F_{B}={\frac{\kappa}{8}}\int (\partial_{xx} z+
\partial_{yy} z)^{2} dx dy
\end{equation}
where $z(x, y)$ denotes small deviation of the plane along the normal direction $(0,0,1)$. Obviously, Eq. (4) 
 is positive for any
violations of a plane since $\kappa$ is always positive. Note that $\delta^{(2)}F_{B}$ is irrelevant to $\kappa_{2}$
and $\kappa_{4}$, which indicates the stability of the plane is irrelevant to the elastic energy terms higher than quadratic 
order in the principal curvatures. To demonstrate this point more generally, 
 we also analyzed the following  curvature elasticity model which includes the 
complete terms up to the fourth order in the principle curvatures
\begin{equation}
F_{B}=\int [ \kappa H^{2}+\overline{\kappa} K+
\kappa_{3} H^{3}+\overline{\kappa}_{3}H K+\kappa_{4} 
H^{4}+\overline{\kappa}_{4} H^{2} K+\kappa_{4}^{*} K^{2} ] dA
\end{equation}
and found the plane is still a strict solution of the corresponding 
shape equation.  However,  the second variation of the energy
 at the plane is found to be the same as  Eq. (4), $i.e.$, irrelevant with 
higher order energy terms.
The second variation $\delta^{(2)} F_{B} >0$
indicates the  existence of an energy barrier between the flat configuration and
any possible stable curved ones, which is in accordance with the result of 
Goetz and Helfrich[15] in their simulations. The deformation of
a plane to a PCST by freezing as observed in Ref. [8-10] thus can not be explained by  considering only curvature elastic energy. Considering the
above mentioned difficulty, in the following we will investigate this problem from
another viewpoint, in which the temperature decreasing  plays a key role in the instability
of the planar configuration.

Generally, when the temperature of the BM is decreased from a high value ($T_h$) to a 
low value ($T_l$), the values of the
 volume Gibbs 
free energy density and the surface tension  will change correspondingly 
because lipid molecules will arrange themselves more
orderly in the  membrane. Thus,  the conversion of the membrane from a high
temperature  state to a low temperature state during the quench-like process is
quite similar to the tube  
formation of
a  smectic-A phase  grown from  isotropic phase in liquid crystal[16, 18] and  the coil formation  of a multishell carbon nanotubes grown from the
carbonaceous mesophase[19]. As shown in Refs. [16, 18, 19],
if take the high temperature state of the membranes as the 
zero energy state, the 
shape formation energy at the low temperature  is the 
sum of the following three terms, 
(i) the net difference of the {\it volume}
  Gibbs free energy 
between the $T_l$ and $T_h$ states, $i.e.$,  $F_{V}=-g_{0} V =-g_{0} d_{0} A$ where
 $V$, $A$, and  $d_{0}$ are  the volume, area, and thickness of the BM, 
respectively, 
 and $g_{0}$ (of positive value) is the difference of the  volume Gibbs free
  energy densities
between the $T_h$ and $T_l$ states, (ii) the 
surface tension energy difference $F_{A}=2(\gamma(T_l)-\gamma(T_h)) A$ where 
$\gamma(T)$ is the surface 
tension  at temperature $T$,  and (iii) the  Helfrich curvature 
elastic energy $F_{B}$ . For symmetric BMs,
 $F_{B}= (\kappa/2)\int H^{2} dA + \overline{\kappa}\int K dA$[1],
(in the following we will also omit the $\overline{\kappa}$ term).Here we
do not take account of higher order elastic terms, because  this
simplest curvature elastic model is enough for following calculations.
 The total shape formation energy is  then
\begin{eqnarray}
F=F_{V}+F_{A}+F_{B}=&&-g_{0} d_{0} \int dA + 
2(\gamma(T_l)-\gamma(T_h)) \int dA + 
{\frac{\kappa}{2}} \int H^{2} dA \nonumber \\
=&& \lambda \int dA +{\frac{\kappa}{2}} \int H^{2} dA ,
\end{eqnarray}
where $\lambda= -g_{0} d_{0}+2 ( \gamma( T_l )-\gamma( T_h ) )$. 

 The variational equation 
of $\delta F=0$ yields the
equilibrium-shape equations of the BM[16, 17],  
\begin{equation}
  \kappa H^{3}-\kappa H K +{\frac{\kappa}{2}} {\nabla}^2 H -2 \lambda H = 0.
\end{equation}
It is obvious that a
planar BM is always a solution of the shape equation (7), 
 and the corresponding shape 
formation energy of the planar BM is $F=\lambda A$.
The shape formation energy is regarded as a free energy  and
the equilibrium threshold condition of
$F=0$ yields the criteria for the stability of a planar BM with the change of temperature
from $T_h$ to $ T_l$ as
\begin{equation}
\lambda=-g_{0} d_{0} +2(\gamma(T_l)-\gamma(T_h)) =0.
\end{equation}
This equation describes the threshold relationship between
 $\gamma$ and $g_{0}$ for a 
planar BM. We emphasize that both $\gamma$ and $g_0$
are dependent on the temperatures and environment conditions.
The deformation procedure for BMs, which is 
 a sudden cooling, is actually a sort of quench-like process.
As long as the shape formation energy for the
planar BMs deviates downwards from the threshold
condition, $i. e.$, $\lambda$ becomes negative, the resultant
remnant part of energy will prevent the  planar BM from keeping stable.
Then a shape deformation will be induced and it would lead to another solution
of the shape equation with its shape formation energy again being equal to zero.
In fact, $g_{0}$ will
increase with the temperature decreasing; moreover, it is argued in
 Ref. [20] 
that when temperature decreases, hydrocarbon chains of lipid molecules
will become less flexible and correspondingly the thickness $d_{0}$ will 
increase slightly. Considering these two effects, according to  Eq. (8), 
$\lambda$ may become negative and the
 planar BM  may  be curved under
the (quick) cooling process.
These features give a natural explanation for the
deformation of BMs. 

Generally, to find  exact periodic  solutions to Eq. (7) is quite
difficult. However, to find a threshold deformation from planar solution
 to the PCST  observed in experiments[8-10] we need only to
consider a  small deformation away 
from  a planar configuration and try to find some approximate 
periodic solutions to Eq. (7). 
The deformed surface is described by Eq. (2). Keeping up to the first order in $z(x,y)$, Eq. (7) reduces to
\begin{equation}
\kappa \Delta H- 4 \lambda H=0
\end{equation}
with
\begin{equation}
 H =(1/2)\Delta z,
 \end{equation}
 where
 $\Delta=\partial_{xx}+\partial_{yy}$ is  the usual two dimensional Laplace operator.
We seek PCST solutions to Eq. (9), which becomes possible
as long as  the value of $\lambda$ changes from positive to negative values.
One such solution  is
\begin{equation}
{\bf Y}(x,y)=(x,y,z_{a}[\cos(2\pi (x-x_{0})/p)+\cos(2\pi (y-y_{0})/p)] ),
\end{equation}
where $x_{0}$ and $y_{0}$ are integration constants, and $p$ is the period,
$p=2\pi \sqrt{\kappa/(-4 \lambda)}$, $z_{a}$ is the amplitude of the deformation. A schematic
representation of the form of this configuration is shown in Fig.~ 1, the 
resemblance of Fig. 1 with experimental observations[8-10] is obvious. 
Based on this agreement, we have confidence to believe that 
the configuration represented by Eq. (11) does reveal
 some essential properties of PCSTs. In the 
following we will discuss some details.

The metric of this PCST surface is
\begin{eqnarray}
\sqrt{g}=\sqrt{1+\partial_{x}^{2}z+\partial_{y}^{2}z}=&&\sqrt{
  1+4 \pi^{2}z_{a}^{2}/p^2} (1-{\frac{2 \pi^{2}z_{a}^{2}[\cos(2 p (x-x_{0}))+
  \cos (2 p (y-y_{0}))]}{p^{2}+4\pi^{2} z_{a}^{2}}})^{1/2} \nonumber \\
  \approx \sqrt{1+4 \pi^{2}z_{a}^{2}/p^{2}}.
\end{eqnarray}
The total surface area of the BM is regarded as conserved, so Eq. (12)
 indicates that the total projected area $A_{p}$  of this PCST  on the 
 $xy$ plane is less than the total area $A$ of the undeformed planar
 configuration, $A_{p}/A=1/\sqrt{1+4 \pi^{2}z_{a}^{2}/p^{2}}=p/\sqrt{
 p^{2}+4\pi^{2}z_{a}^{2}}$. Considering Eqs. (10-12) and the conservation
 of total surface area, the formation energy of this periodic BM can be
 derived from Eq. (6) as
 \begin{equation}
 F=\lambda A (1-{\frac{2\pi^{2}z_{a}^{2}}{p^{2}}})
 \end{equation}
Since now we have the periodic curved situation which is 
different from the planar BM case,  we may treat the threshold condition $F=0$ as 
\begin{equation}
{\frac{p}{z_{a}}}=\sqrt{2} \pi.
\end{equation}
Eq. (14) gives the optimal ratio of the period and the 
amplitude for the PCSTs. Under this condition, the ratio $A_{p}/A=1/\sqrt{3} \approx 0.6$,
$i.e.$, the area projection on the $xy$ plane will cover only about $60$ percent
of the total area occupied by a undeformed planar surface. Of course,
when the amplitude $z_{a}$ becomes comparable with the period, (as is the case
in  Eq. (14) ),
the surface described by Eq. (11) can no longer satisfy the equilibrium-shape
equation (7), so it is just a rough approximation of actual stable PCSTs 
observed in experiments. Even though, we find that the relationship of Eq. (14)
holds well for actual systems,  indicating the validity of the proposed
mechanism for PCST formation.
As shown in Fig. 3b of
Ref. [9], there is a fraction of BMs being regularly curved and the amplitude
of this deformed texture can be estimated from the boundary curve of this 
region. By a direct
evaluation from this figure, we do find  $p/z_{a} \approx \sqrt{2} \pi$ hold 
well. A rough estimation from the periodic deformed part of this figure gives 
$p \approx 25$ nm and $z_{a} \approx 5$ nm, $i.e.$, $ p/z_{a}=5 \approx \sqrt{2}
\pi$. Another
square texture, shown in Fig. 4d of Ref. [8], gives $ p/z_{a} \approx 3/0.5=6$,
again close to the prediction of Eq. (14).

It is also interesting to study the value of $g_{0}$ for the observed PCSTs of the
recent experiments[8-10].
We set $d_{0}=4$ nm[15, 21], $\kappa=4\times 10^{-19}$ J[20]. As for the surface
tension, according to de Gennes and Taupin[22], $\gamma=2 \kappa /(R R_{0})$ where
$R_{0}$ is the  spontaneous radius of curvature and $R$ is the actual radius
of curvature, 
 so  $\gamma( T_{h})=0$[22]
for the planar shape at $T_{h}$, because $R=\infty$, and for 
the PCST shape at $T_{l}$, as an estimation we can set $R \approx R_{0} \approx
p$, hence 
$\gamma( T_{l} ) \approx 2\kappa/p^{2}$.
Then for a typical PCST with period $p=45$ nm
to occur, $g_{0}$ should equal to about $6\times 10^{5}$ J/m$^{3}$. This value
is consistent with the experimental conclusion of Sternberg {\it et al} [9] that
the formation of PCSTs is not connected with temperature-induced lipid
phase transition processes. For if some phase transition should occur, some
thermal energy at least about $1.5-2$ kcal/mol would be released at the transition
point alone, according to Ref. [21]. For a typical BM where each lipid molecule occupies
an average volume of about $1000 \AA^{3}$ [21], it is equivalent to say that there is
at least  a thermal energy difference about $1 \times 10^{7}$ J/m$^{3}$ exists between  the
high and the low temperature states, if a phase transition does occur.
However, the value we estimated from experiments, about $6 \times 10^{5}$ J/m$^{3}$, is
much smaller than this value.
So it is shown that there is no phase transition processes
in the PCST formation  with temperature decreasing,
 which is in agreement with the observation of Ref. [9].

In summary, by deriving the expression for the 
formation energy of the BMs as well as the equilibrium-shape equation, we 
 have shown that there exists a threshold condition for a planar
BM to be stable, below which the planar BMs become unstable and 
will undergo a 
shape deformation. In particular, we have shown the existence of periodic
 deformations  
and given an estimate of the optimal ratio of the period 
and the amplitude, i.e.,
$p/z_{a}\approx \sqrt{2}\pi$,  which is
 consistent with the recent experiment observations[8-10].

This work is partly supported by the National Natural Science Foundation of 
China.

\begin{figure}
\caption{Schematic  representation of the periodic curved square texture 
solution. The ratio of period and amplitude is set equal to $\sqrt{2}\pi$ (see
text for details).  }
\end{figure}

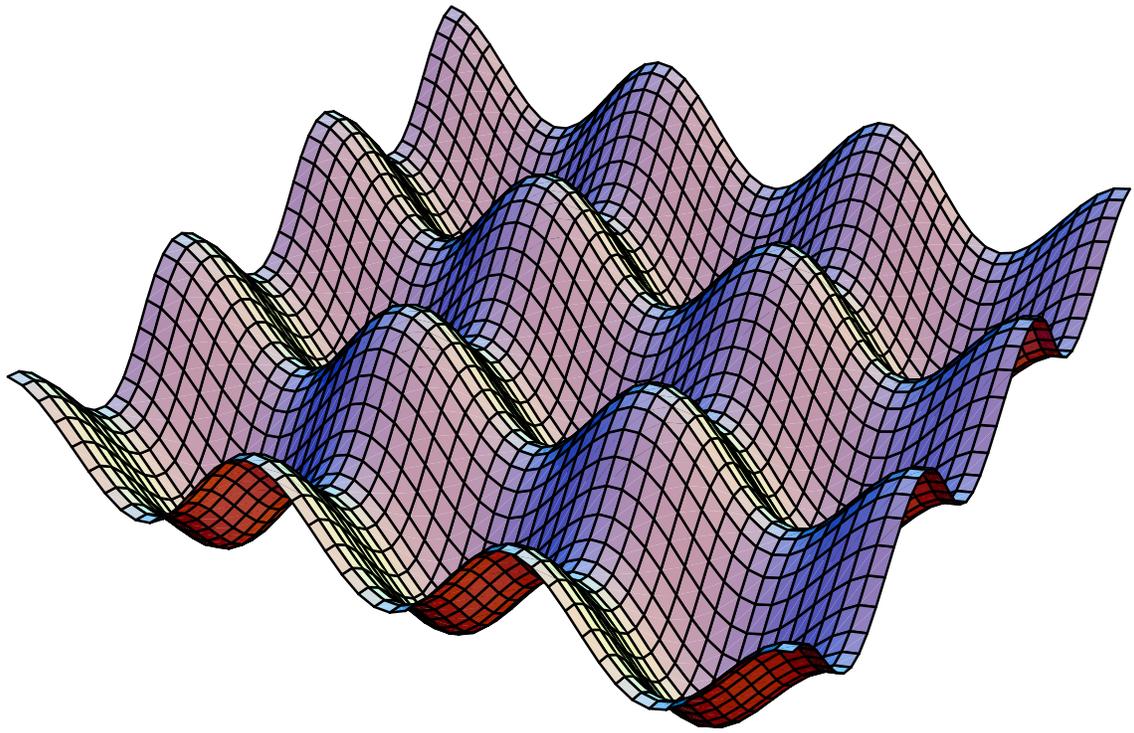
\end{document}